\documentstyle [12pt]{article}

\begin{document}

\hoffset = -1truecm
\voffset = -3truecm

\author{Mikhail S. Plyushchay\thanks{Present address:
Departamento de F\'{\i}sica Te\'orica, Universidad de Zaragoza, 50009 Zaragoza,
Spain; e-mail: mikhail@dedalo.unizar.es} {}\thanks{On leave from
the Institute for High Energy Physics, Protvino, Russia}\\ \\
{\it International Centre for Theoretical Physics, I-34100 Trieste,
Italy}}
\date{{\bf Mod. Phys. Lett. A10 (1995) 1463-1469}}

\title{{\bf RELATIVISTIC PARTICLE WITH TORSION AND
CHARGED PARTICLE IN A CONSTANT ELECTROMAGNETIC
FIELD:\\ IDENTITY OF EVOLUTION}}

\date{{\bf Mod. Phys. Lett. A10 (1995) 1463-1469}}

\maketitle

\begin{abstract}
The identity of classical motion is established for two
physically different models, one of which is the relativistic
particle with torsion, whose action contains higher derivatives
and which is the effective system for the statistically charged
particle interacting  with the Chern-Simons U(1) gauge field,
and another is the (2+1)-dimensional relativistic charged
particle in external constant electromagnetic field.
\end{abstract}

\vfill

\newpage
The models of relativistic particles and strings, which are
described by
lagrangians containing the dependence on
higher (over evolution parameter)
derivatives of the space-time coordinates $x^\mu$,
have a very nontrivial, from the point of
view of ordinary relativistic systems, property:
the sign of their squared energy-momentum vectors generally
turns out to be not correlated with the sign of the corresponding
squared velocity vectors.
For example,
massless and tachyonic solutions can take place in such systems
in the case of their motion with
the velocities less than the velocity of
light \cite{rigm}--\cite{aro}. Moreover, one of the known models,
the so called
massless particle with rigidity \cite{rig0},
classically is unambiguous only in the case of the
superrelativistic motion of the particle, but
corresponding solutions to the equations of motion of the system
are given by the isotropic energy-momentum vector.

Despite this very unusual property, such models are physically
interesting systems.
The first model of the type,
relativistic string with rigidity,
was proposed as a QCD-inspired string model \cite{polstr}, and it was
also considered in the context of the condensed
matter physics \cite{kle}.
The massive particle with rigidity \cite{rigm},
related to the infinite-component Majorana equation \cite{maj},
finds the application for explaining the nature of narrow
$e^{+}e^{-}$ peaks resulting from heavy ion collisions \cite{awad},
whereas the above-mentioned massless particle with rigidity,
being a nontrivial model for  the massless bosons and fermions
\cite{rig0bf},
turns out also to be related to W-algebras \cite{rig0w}.

We shall reveal here the relationship of one such higher
derivative model, the model of relativistic particle with
torsion \cite{tor}, to the ordinary relativistic system without
higher derivatives, which is a charged relativistic massive point
particle interacting with a specific external electromagnetic
field. This will allow us to get a natural explanation for
the above-mentioned nontrivial property
for higher derivative systems in the particular case
of the system under consideration.
Moreover, it seems that such a relationship could be useful for
investigation and understanding general nature and properties
of the higher derivative systems.

The action of the relativistic particle with torsion,
\begin{equation}
S_{tor}=-\int(m+\alpha\kappa)ds,\quad
\kappa=k^{-2}\epsilon^{\mu\nu\lambda}x'_{\mu}x''_{\nu}x'''_{\lambda},
\quad k^{2}=x''{}^{2},
\label{act}
\end{equation}
appeared under consideration of describing high-$T_{c}$
superconductivity \cite{pol}. It was evaluated within
a path-integral approach as an effective action for the
(2+1)-dimensional relativistic charged scalar massive particle
interacting with a U(1) gauge field having the Chern-Simons kinetic term
\begin{equation}
{\cal L}_{CS}=\frac{\vartheta}{4\pi^{2}}\epsilon^{\mu\nu\lambda}A_{\mu}
\partial_{\nu}A_{\lambda}.
\label{cs}
\end{equation}
The term with torsion $\kappa$ in the effective action (\ref{act})
is induced by the Chern-Simons gauge field, and the parameter $\alpha$
is connected with the constant $\vartheta$ and the charge
of the particle $q_{s}$ through the relation
$\alpha=\pi q_{s}^{2}/2\vartheta$ \cite{pol,tze}, where
$q_{s}$ has a sense of a statistical charge.
In eq. (\ref{act}) we use the notations
$ds^{2}=-dx^{\mu}dx^{\nu}\eta_{\mu\nu},$ $\eta_{\mu\nu}={\rm
diag}(-1,1,1)$, $x'_{\mu}=dx_{\mu}/ds$, and assume that
$\epsilon^{012}=+1$ for the totally antisymmetric tensor
$\epsilon^{\mu\nu\lambda}$.

The classical theory of system (\ref{act}) was considered in
details in refs. \cite{tor,torc}.  There it was also shown that
the quantization of the model leads to the (2+1)-dimensional
analog of the Majorana equation \cite{maj} being a basic
equation for the description of the fractional spin fields
within a framework of the group-theoretical approach to anyons
\cite{tor},\cite{jn}--\cite{corp}.

In the present letter we shall demonstrate that the equations of
motion, following from action (\ref{act}), which are the third
order differential equations for the coordinates of the particle
over the evolution parameter, coincide with the form of
the equations of
motion for the (2+1)-dimensional relativistic electrically
charged massive particle interacting with the external constant
homogeneous electric and magnetic fields.
Moreover, via the identification
(up to the constant factor)
of the energy-momentum vector of
the system (\ref{act}) with the (2+1)-dimensional vector dual
to the constant electromagnetic field tensor for the second system,
we shall reveal the identity of evolution for
the particles in these two physically different systems.

To begin with, we rewrite action
(\ref{act}) in the given parametrization $x_{\mu}=x_{\mu}(\tau)$:
\begin{equation}
S_{tor}=\int Ld\tau,\quad
L=-\sqrt{-\dot{x}{}^{2}}\left(m-\alpha
\frac{\epsilon^{\mu\nu\lambda}\dot{x}_{\mu}
\ddot{x}_{\nu}\stackrel{...}{x}{}_{\lambda}}{\ddot{x}{}^{2}\dot{x}{}^{2}-
(\ddot{x}\dot{x})^{2}}\right).
\label{act2}
\end{equation}
The Lagrange equations of motion, following from action
(\ref{act2}), have the form \cite{tor}:
\begin{equation}
\frac{dp^{\mu}}{d\tau}=0,
\label{eqt}
\end{equation}
where
\begin{equation}
p^{\mu}=me^{\mu}+\frac{\alpha}{\sqrt{-\dot{x}{}^{2}}}
\epsilon^{\mu\nu\lambda}e_{\nu}\dot{e}_{\lambda}
\label{pmu}
\end{equation}
is the energy-momentum vector of the system,
and $e^{\mu}=\dot{x}{}^{\mu}/\sqrt{-\dot{x}{}^{2}}$ is the relativistic
three-velocity of the particle. Third order differential
equations (\ref{eqt})
are equivalent to the system of the second and first order equations
\begin{equation}
\dot{e}^{\mu}=\alpha^{-1}\sqrt{-\dot{x}{}^{2}}
\epsilon^{\mu\nu\lambda}p_{\nu}e_{\lambda},\qquad
\dot{p}{}^{\mu}=0
\label{eqt2}
\end{equation}
with the additional condition
\begin{equation}
pe+m=0.
\label{con}
\end{equation}
Eqs. (\ref{eqt2}) are in fact the Hamiltonian equations, whereas condition
(\ref{con}) is a Hamiltonian constraint, which presents by
itself the restriction on the initial data of the system  \cite{tor}.
{}From eq. (\ref{pmu})  follows that
the squared mass of the system, $M^{2}=-p^{2}$,
takes positive, zero and negative values when the squared curvature
of the world particle trajectory,
$
k^{2}=(\ddot{x}{}^{2}\dot{x}{}^{2}-(\ddot{x}\dot{x})^{2})/
(\dot{x}{}^{2})^{3},
$
takes the values
$
0\leq k^{2}<m^{2}\alpha^{-2},$
$k^{2}=m^{2}\alpha^{-2}$ and
$k^{2}>m^{2}\alpha^{-2}$, respectively.

Therefore, though we have assumed that the particle has the velocity less
than the velocity of light, $\dot{x}{}^{2}<0$,
there are massive, massless and tachyonic
sectors in the system (see below).

Now, let us consider (2+1)-dimensional relativistic
electrically charged particle, interacting with the external
electromagnetic field described by the electromagnetic potential
$A_{\mu}=A_{\mu}(x)$. This system is given by the action:
\begin{equation}
S_{em}=\int_{}^{} d\tau(-m\sqrt{-\dot{x}{}^{2}}+q_{em}
A_{\mu}\dot{x}{}^{\mu}),
\label{em}
\end{equation}
which is also reparametrization invariant similarly
to action (\ref{act2}).
Here $q_{em}$ is the electric charge of the particle and
for the sake of simplicity we take the same mass parameter as in the
previous system.
The equations of motion following from  action (\ref{em}) have the form
\begin{equation}
m\dot{e}^{\mu}=q_{em}\sqrt{-\dot{x}{}^{2}}\epsilon^{\mu\nu\lambda}R_{\nu}
e_{\lambda},
\label{eqem}
\end{equation}
where (2+1)-dimensional vector $R_{\mu}$ is dual to the
electromagnetic field tensor
$F_{\mu\nu}=\partial_{\mu}A_{\nu}-\partial_{\nu}A_{\mu}$:
$R^{\mu}=\frac{1}{2}\epsilon^{\mu\nu\lambda}F_{\nu\lambda}$, and
has the components
\begin{equation}
R^{\mu}=(H, E^{2},-E^{1})
\label{rmu}
\end{equation}
with $E^{i}=F^{0i}$, $i=1,2,$
and $H=F^{12}$ being the electric and magnetic fields.
Eqs. (\ref{eqem}) will coincide in the form with
eqs. (\ref{eqt2}) if we suppose that
$R^\mu$ does not depend on $x_\mu$ and
\[
\frac{d}{d\tau}R^{\mu}=0,
\]
i.e. assume that the external electric and magnetic fields are
homogeneous constant fields,
and then make the formal identification:
\begin{equation}
p^{\mu}=\alpha q_{em}m^{-1} R^{\mu}.
\label{ide}
\end{equation}
As a result,
we have the integral of motion
\begin{equation}
I=R^{\mu}e_{\mu}
\label{mi}
\end{equation}
for the system under consideration,
which, e.g., in the laboratory temporal gauge, $x^{0}\equiv t=\tau$,
takes the form:
\[
I=\frac{E^{2}v^{1}-E^{1}v^{2}-H}{\sqrt{1-v^{i}v^{i}}},
\]
where $v^{i}=dx^{i}/dt$.
Therefore, if in correspondence with eq. (\ref{con}),
this integral of motion takes the value
\begin{equation}
I=-\alpha^{-1} q_{em}^{-1}m^{2},
\label{con2}
\end{equation}
we shall have the identical evolution for the both systems.

In the proper time gauge, $\dot{x}{}^{2}=-1$,
the equations of motion (\ref{eqt2}) in the
massive sector have the solutions:
\begin{eqnarray}
x_{\mu}(\tau)&=&x_{\mu}(0)+\frac{\alpha}{M}e_{\mu}(0)\sin\omega\tau-
\frac{\alpha}{M^{2}}\epsilon_{\mu\nu\lambda}p^{\nu}e^{\lambda}(0)(\cos
\omega\tau-1)+
\nonumber\\& &+\alpha\frac{pe(0)}{M^{3}}p_{\mu}(\sin\omega\tau-\omega
\tau),
\label{massiv}
\end{eqnarray}
where $M=\sqrt{-p^{2}},$ $\omega=\alpha^{-1}M,$
$p_{\mu}=p_{\mu}(0)=const$.
In the massless sector, $p^{2}=0$, the solutions are
\begin{equation}
x_{\mu}(\tau)=x_{\mu}(0)+e_{\mu}(0)\tau+
\alpha^{-1}\epsilon_{\mu\nu\lambda}p^{\nu}e^{\lambda}(0)\frac{\tau^{2}}{2}
-\alpha^{-2}(pe(0))p_{\mu}\frac{\tau^{3}}{6},
\label{massl}
\end{equation}
whereas in the tachyonic sector, $p^{2}>0$, the form of the solutions is
analogous to (\ref{massiv}), the trigonometric functions being substituted
by the corresponding hyperbolic functions:
\begin{eqnarray}
x_{\mu}(\tau)&=&x_{\mu}(0)+\frac{\alpha}{\mu}e_{\mu}(0)\sinh\tilde\omega
\tau+
\frac{\alpha}{\mu^{2}}\epsilon_{\mu\nu\lambda}p^{\nu}e^{\lambda}(0)(\cosh
\tilde\omega\tau-1)-\nonumber\\& &-\alpha\frac{pe(0)}{\mu^{3}}p_{\mu}
(\sinh\tilde\omega\tau-\tilde\omega\tau)
\label{tach}
\end{eqnarray}
with $\mu=\sqrt{p^{2}},$ $\tilde\omega=\alpha^{-1}\mu$.
In all the three cases, in correspondence with the above-said, the vectors
$p_{\mu}$ and $e_{\mu}(0)$ obey the condition $pe(0)+m=0$
as a condition imposed on the initial data of the first system.
The explicit form of eqs. (\ref{massiv})--(\ref{tach})
implies that the massless solutions
(\ref{massl}) can be obtained from the solutions in the massive and tachyonic
sectors via the corresponding limit transition:  $M\rightarrow0$ or
$\mu\rightarrow0$. At the same time, the solutions in the tachyonic sector
(\ref{tach}) can, in turn,  be obtained from those in
the massive sector by changing  $M\rightarrow i\mu$ in (\ref{massiv}).
Eqs. (\ref{massiv}), (\ref{massl}) and (\ref{tach}) together with eq.
(\ref{ide}) also give the solutions to the equations of motion of the second
system for the cases
$R^{2}=(E^{1})^{2}+(E^{2})^{2}-H^{2}<0$,
$R^{2}=0$ and $R^{2}>0$, respectively.

Now, let us make two remarks.
First, it is obvious that the described (2+1)-dimensional system
of the charged particle in external constant electromagnetic
field can be considered as a (3+1)-dimensional charged particle,
confined in the plane $x^{3}=0$ and interacting with external
crossing constant electric and magnetic fields ${\bf E}=
(E^{1},E^{2},0)$ and ${\bf H}=(0,0,H)$.
In this case  corresponding electromagnetic field tensor has components
$F_{\mu 3}=F_{3\mu}=0$.
Note, that such a system is realized in
quantum Hall effect experiments.

The second remark
concerns the properties of the two systems with respect
to the $P-$ and $T-$inversions:
\begin{equation}
P:\,x^{\mu}\rightarrow (x^{0},-x^{1},x^{2}),\quad
T:\,x^{\mu}\rightarrow (-x^{0},x^{1},x^{2}).
\label{pt}
\end{equation}
The model of relativistic particle with torsion is $P-$ and
$T-$noninvariant (as well as the initial relativistic particle interacting
with the Chern-Simons gauge field).
But it is invariant with respect to each of these
operations supplemented with the operation of the change of the parameter
sign: $\alpha\rightarrow -\alpha$, and, therefore, is invariant with
respect to the $PT-$inversion. The system of (2+1)-dimensional
charged particle interacting with electromagnetic field,
described by the action (\ref{em}),
is invariant with respect to $P-$ and $T-$inversions (\ref{pt})
supplemented with the relations
$
P:\,R^{\mu}\rightarrow (-R^{0},R^{1},-R^{2}),$
$T:\,R^{\mu}\rightarrow (R^{0},-R^{1},-R^{2}).$
But since
$
P:\,Re\rightarrow -Re,$
$T:\,Re\rightarrow -Re,$
the additional condition (\ref{con2}) violates these two symmetries,
preserving the
combined $PT-$symmetry. From the above-said it follows that
for the first system,
$P-$ and $T-$inversions
transform the solutions with $\alpha>0$ into the solutions
with $\alpha<0$ and vice versa.
For the second system, these operations transform mutually
the solutions with $I>0$ and $I<0$.
Therefore, generally
the second system has $P-$ and $T-$invariant solutions corresponding to
$I=0$, which due to equality $e^{2}=-1$,
take place only in the sector with $R^{2}>0$, and their analogs
are not contained  in the first system.

So, we have proved the identity of the evolution of
the particle with torsion (\ref{act}), which is the effective
system corresponding to the statistically charged particle
interacting with the Chern-Simons gauge field, and the
(2+1)-dimensional electrically charged particle interacting with
the external constant electromagnetic field.  Such a phenomenon
takes place under identification (\ref{ide}) between the
energy-momentum vector of the particle with torsion and the
(2+1)-dimensional vector (\ref{rmu}) constructed from the
constant electric and magnetic fields for the case of the
electrically charged particle.
In other words, we have shown that every solution of equations of motion
for the first system, which is given by the conserved energy-momentum
vector $p^{\mu}$  (and by the initial data $x^{\mu}(0)$ and $e^{\mu}(0)$),
corresponds exactly to the solution of equations of motion for
the relativistic charged particle moving in constant homogeneous
electric and magnetic
fields with the values $E^{1}=-cp^{2}$, $E^{2}=cp^{1}$, $H=cp^{0}$,
where $c=m\alpha^{-1}q^{-1}_{em}$.
At the same time, this identity
does not mean the complete equivalence of the
described two systems since the vector $p^{\mu}$ is the internal
energy-momentum vector for the first system, whereas the vector
$R^{\mu}$ is the external vector for the electrically charged
particle.  Besides, condition (\ref{con}) is the internal one
for the first system, but condition (\ref{con2}) is the
additional condition fixing the value of the integral of motion
(\ref{mi}) for the second system, and as we have seen,
there is a class of solutions for system (\ref{em}) with
$I=Re=0$, $R^{2}>0$, $R^{\mu}=const$,
whose analog is not contained among the solutions
for system (\ref{act2}).

Thus, we conclude that the system with higher derivatives,
the model of relativistic particle with torsion, can be
modelled by the ordinary system without higher derivatives.
This ordinary system
is relativistic charged particle interacting with the
specific external electromagnetic field, given by the constant
homogeneous electric, ${\bf E}$, and magnetic, $H$, fields.
 From this point of view we get a natural explanation for the
existence of massive, massless and tachyonic solutions
with timelike velocities of the particle in the first system.
These three types of solutions
correspond simply to three possible ``configurations"
of the electromagnetic field in the
second system, which are characterized by the timelike,
$R^2={\bf E}^2-H^2<0$,
isotropic, $R^2=0$, and spacelike, $R^2>0$, vector $R^\mu=(H,E^2,-E^1)$
dual to the electromagnetic field tensor.

Finally, we arrive at
a natural hypothesis that other
higher derivative systems such as massive \cite{rigm} and
massless \cite{rig0} particles with rigidity and relativistic
string  with rigidity \cite{polstr} can be modelled by some
ordinary systems (without higher derivatives)
of relativistic particles and strings interacting with
some specific external fields.
Hence, it seems to be interesting to
continue investigation of the revealed
classical relationship, and,
moreover, analyse the possible consequences of such a relationship
from the point of view of the
quantum theory for the higher derivative systems.

\vskip \baselineskip

{\bf Acknowledgements.}

The author thanks Prof. Abdus Salam, the International Atomic Agency and
${\rm UNESCO}$ for hospitality at
the International Centre for Theoretical Physics, Trieste.
He is also grateful to I.A. Bandos,
D.P. Sorokin and D.V. Volkov for interesting discussions
and to M. Tonin and P.-A. Marchetti
for discussions and hospitality
at Dipartimento di Fisica, Universit\'a di Padova.

The research was supported, in part, by a Soros Foundation Grant awarded
by the American Physical Society.

\end{document}